\documentclass[final]{elsart}

\usepackage[dvips]{graphicx}

\usepackage{amssymb}
\usepackage{amsmath}

\begin{document}

\begin{frontmatter}

\title{Effect of multiple reusing of simulated air showers in detector simulations}

\author[UNAM]{A. D. Supanitsky\corauthref{cor}} and
\author[UNAM]{G. Medina-Tanco}
\address[UNAM]{Instituto de Ciencias Nucleares, UNAM, Circuito Exteriror S/N, Ciudad Universitaria,
M\'exico D. F. 04510, M\'exico.}
\corauth[cor]{Corresponding author. E-mail: supanitsky@nucleares.unam.mx.}

\begin{abstract}
The study of high energy cosmic rays requires detailed Monte Carlo simulations of both, extensive air 
showers and the detectors involved in their detection. In particular, the energy calibration of several 
experiments is obtained from simulations. Also, in composition studies simulations play a fundamental role 
because the primary mass is determined by comparing experimental with simulated data. At the highest energies 
the detailed simulation of air showers is very costly in processing time and disk space due to the large number 
of secondary particles generated in interactions with the atmosphere. Therefore, in order to increase the 
statistics, it is quite common to recycle single showers many times to simulate the detector response. As a 
result, the events of the Monte Carlo samples generated in this way are not fully independent. In this work 
we study the artificial effects introduced by the multiple use of single air showers for the detector simulations. 
In particular, we study in detail the effects introduced by the repetitions in the kernel density estimators which 
are frequently used in composition studies.  
\end{abstract}

\begin{keyword}
Cosmic Rays, Air Showers, Detector Simulations
\end{keyword}
\end{frontmatter}

\section{Introduction}
\label{int}

The spectrum of cosmic rays extends over more than eleven orders of magnitude, starting at $E\sim 10^9$ eV
up to energies above $10^{20}$ eV. Above $\sim 10^{14}$ eV they are too infrequently to be detected by 
balloons or spacecraft. Therefore, the detection techniques used in this energy range are based in the 
properties of the extensive air showers produced by them in the atmosphere. There are essentially 
two techniques for shower detection \cite{Alan:00}: ($i$) arrays of surface detectors 
which measure the lateral distribution of the secondary particles that reach the Earth surface and ($ii$) 
measurements of the fluorescence light emitted by atmospheric nitrogen excited by charged 
particles of the shower as they traverse the atmosphere.    

Air shower and detector simulations play a fundamental role in the study of cosmic rays. In particular,
arrays of surface detectors that do not have fluorescence telescopes to calibrate the energy scale, 
must resort to simulated data in order to estimate the energy of the primary particle. Furthermore, the 
primary mass is also obtained comparing experimental data with simulations.

There are several Monte Carlo programs for air shower simulation, the most used in the literature are 
AIRES \cite{AIRES}, CORSIKA \cite{CORSIKA}, and CONEX \cite{conex}, the latter for a fast simulation of the 
longitudinal shower development. Since the number of particles produced in a shower can be extremely large, e.g., 
$\sim 10^{11}$ for a $10^{20}$ eV proton shower, the computer processing time and disk space needed are also very 
large, even if unthinning methods \cite{Hillas:85,Hillas:97} are used. Due to this difficulty it is a common 
practice to reuse the same shower for generating several events (see for example 
\cite{Ave:01,Ave:02,Dova:03,DovaICRC:03,Vitor:05,VitorFede:05}). This practice is more common in simulations 
that includes surface detectors because, for fluorescence telescopes, very fast Monte Carlo programs like 
CONEX, introduced few years ago, have very fast and efficient algorithms for the generation of longitudinal 
profiles. 

In this work we study the effects of using multiple repetitions of individual showers, applied to the simulation of  
detectors, on the evaluation of standard estimators of the expected value, variance, and covariance as well as on 
histograms corresponding to observable parameters. We study in detail the effects introduced in the kernel density 
estimators, which are analytical estimates of the underlying distribution function obtained from a finite sample of 
events. In cosmic rays physics this technique is used mainly in connection with composition analyses 
\cite{Rebel:95,Brancus:97,Antoni:07,Antoni:03,Brancus:03,SupaCompo:08}; however, it is also extensively 
used in many different areas of knowledge \cite{Silvermann:86} to which this work can be directly extended. 

As a numerical example, we discuss the effects of repetitions on samples of the $X_{max}$ parameter, the atmospheric 
depth at which an air shower reach its maximum development, obtained with the package CONEX.

\section{Analytical Treatment}

As mentioned in the introduction, we want to study the potential distortions introduced by reusing individual showers 
to maximize the statistics when simulating the response of a detector. Let us start with the optimum case in which each 
individual shower is used only once and, therefore, best reproduces reality. 

Let $\mathbf{y}$ be a $d$-dimensional vector composed by physical observables (e.g. mass sensitive parameters) 
distributed as $g(\mathbf{y})$ and let $\mathbf{z}$ be a random vector, distributed as $h(\mathbf{z})$, that takes 
into account the effects of the detectors and the corresponding reconstruction method such that, after measuring 
and reconstructing the empirical information, a vector $\mathbf{x}=\mathbf{y}+\mathbf{z}$ is obtained. The distribution 
function of $\mathbf{x}$ is the convolution of $g(\mathbf{y})$ and $h(\mathbf{z})$,
\begin{equation}
f(\mathbf{x}) = g\circ h(\mathbf{x}) = \int d\mathbf{y}\ g(\mathbf{y}) h(\mathbf{x}-\mathbf{y}). 
\label{fdef}
\end{equation}

Suppose that we have a sample of $N$ independent events of the distribution $f$
\begin{eqnarray}
\mathbf{x}_1 &=& \mathbf{y}_1 + \mathbf{z}_1 \nonumber \\
&\vdots& \nonumber \\
\mathbf{x}_N &=& \mathbf{y}_N + \mathbf{z}_N. \nonumber 
\label{samp}
\end{eqnarray}
The probability of this configuration can be written as,
\begin{eqnarray}
P(\mathbf{y}_1 \ldots \mathbf{y}_N, \mathbf{z}_1 \ldots \mathbf{z}_N) &=& g(\mathbf{y}_1)\ldots g(\mathbf{y}_N)\ h(\mathbf{z}_1)%
\ldots h(\mathbf{z}_N), \\ 
P(\mathbf{x}_1 \ldots \mathbf{x}_N) &=& f(\mathbf{x}_1) \ldots f(\mathbf{x}_N).
\label{Dsamp}
\end{eqnarray}

However, as previously noted, if single showers are recycled and used many times to simulate the response of the 
detectors, non-independent samples are obtained. If we use each shower of a sample of $M$ independent showers $m$ 
times to simulate the detectors  response, the following sample of size $N=M\times m$ is obtained, 
\begin{eqnarray}
\mathbf{x}_{11} &=& \mathbf{y}_1 + \mathbf{z}_{11} \nonumber \\
&\vdots& \nonumber \\
\mathbf{x}_{1m} &=& \mathbf{y}_1 + \mathbf{z}_{1m} \nonumber \\
&\vdots& \nonumber \\
\mathbf{x}_{M1} &=& \mathbf{y}_M + \mathbf{z}_{M1} \nonumber \\
&\vdots& \nonumber \\
\mathbf{x}_{Mm} &=& \mathbf{y}_M + \mathbf{z}_{Mm}, \nonumber 
\label{samprep}
\end{eqnarray}
where the notation used henceforth corresponds to $\xi^i_{\alpha a}$, where $i$ is the $ith$ coordinate of vector 
$\mathbf{\xi}$, $\alpha$ indicates the number of independent shower and $a$ the number of detector simulation 
performed using the $\alpha$-$th$ shower. The probability of such a configuration is given by
\begin{eqnarray}
P(\mathbf{y}_1 \ldots \mathbf{y}_N, \mathbf{z}_{11} \ldots \mathbf{z}_{Mm})%
&=& \prod_{\alpha=1}^M g(\mathbf{y}_\alpha) \prod_{a=1}^m h(\mathbf{z}_{\alpha a}) \\
P(\mathbf{x}_{11} \ldots \mathbf{x}_{Mm}) &=& \prod_{\alpha=1}^M \int d\mathbf{y}_\alpha\ g(\mathbf{y}_\alpha)%
\prod_{a=1}^m h(\mathbf{x}_{\alpha a}-\mathbf{y}_{\alpha}).
\label{DsampRep}
\end{eqnarray}

\subsection{Mean, variance and covariance estimators}

Let us consider the average of the $ith$ coordinate of $\mathbf{x}$, $x^i$, for the realistic case
in which each shower is used only once to simulate the detector response,
\begin{equation}
\bar{x}^i = \frac{1}{N}\ \sum_{\alpha=1}^N x_\alpha^i.
\label{xav}
\end{equation}
By using Eq. (\ref{Dsamp}) it is easy to obtain the very well known expressions for the expected value and 
variance of $\bar{x}^i$,
\begin{eqnarray}
\label{Ex}
E[\bar{x}^i]&=&E[x^i] \\
\label{Vx}
Var[\bar{x}^i] &=& \frac{1}{N}\ Var[x^i].
\end{eqnarray}

The usual estimator of the covariance between two random variables is given by,
\begin{equation}
\label{CovEst}
\hat{C}_{ij} = \frac{1}{N-1}\ \sum_{\alpha=1}^N  (x_\alpha^i-\bar{x}^i) (x_\alpha^j-\bar{x}^j).
\end{equation}
For $i=j$ the estimator of the variance of $x^i$ is obtained, $s_i^2=\hat{C}_{ii}$. By using Eq. (\ref{Dsamp}) it can be 
shown that both estimators are non-biased,
\begin{eqnarray}
\label{Ecov}
E[\hat{C}_{ij}]&=&cov[x^i, x^j], \\
\label{Es}
E[s_{i}^2] &=& E[\hat{C}_{ii}]= Var[x^i].
\end{eqnarray}

For the case in which each shower is used several times to simulate the response of the detectors the average of $x^i$ 
is given by,
\begin{equation}
\bar{x}'^i = \frac{1}{M m}\ \sum_{\alpha=1}^M \sum_{a=1}^m x_{\alpha a}^i.
\label{AvRep}
\end{equation}

From Eqs. (\ref{DsampRep},\ref{AvRep}) it can be shown that,
\begin{eqnarray}
\label{Eavxrep}
E[\bar{x}'^i] &=& E[x^i], \\
\label{Vavxrep}
Var[\bar{x}'^i] &=& \frac{1}{Mm}\ Var[x^i]+\frac{m-1}{Mm}\ \int d\mathbf{y} d\mathbf{x}_1 d\mathbf{x}_2%
\ (x^i_1-E[x^i])(x^i_2-E[x^i])\times \nonumber \\
&& g(\mathbf{y})\ h(\mathbf{x}_1-\mathbf{y})\ h(\mathbf{x}_2-\mathbf{y}),
\end{eqnarray}
which means that using samples obtained by reusing individual showers to simulate the detector 
response does not introduce any bias when calculating the average. However the fluctuations of 
$\bar{x}^i$ are increased by the generation of an additional term proportional to $(m-1)/Mm$. 

If the response of the detectors and the reconstruction methods do not introduce any bias on the physical 
magnitudes $\mathbf{y}$, i.e., $\int d\mathbf{u}\ u^i\ h(\mathbf{u}) = 0$, the variance of $\bar{x}^i$
can be written as,
\begin{eqnarray}
\label{Eavxrep}
Var[\bar{x}'^i] &=& \frac{1}{Mm}\ Var[x^i]+\frac{m-1}{Mm}\ Var[y^i], \\
\label{Eavxrep2}
&=& \frac{1}{M}\ Var[y^i]+\frac{m-1}{Mm}\ Var[z^i],
\end{eqnarray}
where $Var[x^i]=Var[y^i]+Var[z^i]$ is used to obtain the last equation.
 
The estimator of the covariance, between $x^i$ and $x^j$, including multiple repetitions of the individual 
showers takes the form,
\begin{equation}
\hat{C}'_{ij} = \frac{1}{M m-1}\ \sum_{\alpha=1}^M \sum_{a=1}^m (x_{\alpha a}^i-\bar{x}'^i)%
(x_{\alpha a}^j-\bar{x}'^j).
\label{CovRep}
\end{equation}

The expected value of the covariance estimator is obtained from Eqs. (\ref{DsampRep}) and (\ref{CovRep}), 
\begin{eqnarray}
\label{Ecovxrep}
E[\hat{C}'_{ij}] &=& cov[x^i, x^j]-\frac{m-1}{Mm} \int d\mathbf{y} d\mathbf{x}_1 d\mathbf{x}_2%
\ (x^i_1-E[x^i])(x^j_2-E[x^j])\times \nonumber \\
&& g(\mathbf{y})\ h(\mathbf{x}_1-\mathbf{y})\ h(\mathbf{x}_2-\mathbf{y}).
\end{eqnarray}
Therefore, as expected, the repetition of individual showers introduces a bias in the covariance estimator
because the events are not independent. The bias results proportional to $(m-1)/Mm$. 

As mentioned before, the expected value of the variance estimator is obtained setting $i=j$ in Eq. 
(\ref{Ecovxrep}),
\begin{eqnarray}
\label{Evarxrep}
E[{s'}_i^2] &=& Var[x^i]-\frac{m-1}{Mm} \int d\mathbf{y} d\mathbf{x}_1 d\mathbf{x}_2%
\ (x^i_1-E[x^i]) (x^i_2-E[x^i])\ g(\mathbf{y})\times \nonumber \\
&& h(\mathbf{x}_1-\mathbf{y})\ h(\mathbf{x}_2-\mathbf{y}),
\end{eqnarray}
which shows that also ${s'}_i^2$ is now a biased estimator of the variance of $x^i$.

For the case in which the detectors and reconstruction methods do not introduce any bias 
Eqs. (\ref{Ecovxrep},\ref{Evarxrep}) become,
\begin{eqnarray}
\label{Ecovxrep2}
E[\hat{C}'_{ij}] &=& cov[x^i, x^j]-\frac{m-1}{Mm}\ cov[y^i, y^j], \\
\label{Evarxrep2}
E[{s'}_i^2] &=& Var[x^i]-\frac{m-1}{Mm}\ Var[y^i].
\end{eqnarray}

\subsection{Histogram fluctuations}

The fluctuations in each bin of a histogram are also modified by the repetition of individual showers. For 
simplicity let us consider a one-dimensional histogram of $N_B$ bins, such that a given $x$ belongs to
the $kth$ bin if $x \in [t_k, t_k+\Delta t]$, where $\Delta t$ is size of the bin.  

The fluctuations in the content $n_k$ of the $kth$-bin of a histogram follows a binomial distribution. Therefore, 
the expected value and the variance of $n_k$ are given by,
\begin{eqnarray}
\label{Ehisn}
E[n_k] &=& N p_k, \\
\label{Varhisn}
Var[n_k] &=& N p_k (1-p_k),
\end{eqnarray}
where 
\begin{equation}
p_k = \int_{t_k}^{t_{k+1}} dx f(x),
\label{Pk}
\end{equation}
with $f(x)=g\circ h(x)$.

The random variable $n_k$ corresponding to a sample of $m$ repetitions of each individual shower can be written as
\begin{equation}
n'_k = \sum_{\alpha =1}^{M}  \sum_{a =1}^{m} \left[ \Theta(x_{\alpha a}-t_k)-\Theta(x_{\alpha a}-t_{k+1}) \right],
\label{nk}
\end{equation}
where $\Theta(x) =1$ if $x\geq 0$ and $\Theta(x) =0$ otherwise. Written in this way it is easy to calculate the 
expected value and variance of $n_k$,
\begin{eqnarray}
\label{Ehisnrep}
E[n'_k] &=& M m\ p_k, \\
Var[n'_k] &=& M m\ p_k (1-p_k) + M m (m-1) \left[ \int dy\ g(y) \left( \int_{t_k}^{t_{k+1}} dx h(x-y) \right)^2 \right. \nonumber \\
\label{Varhisnrep}
&& \left.-\left( \int dy\ g(y) \int_{t_k}^{t_{k+1}} dx h(x-y) \right)^2 \right],
\end{eqnarray}
i.e., the mean value does not change and the variance has an extra term that increases with $m$.

\begin{figure}[!bt]
\begin{center}
\includegraphics[width=10cm]{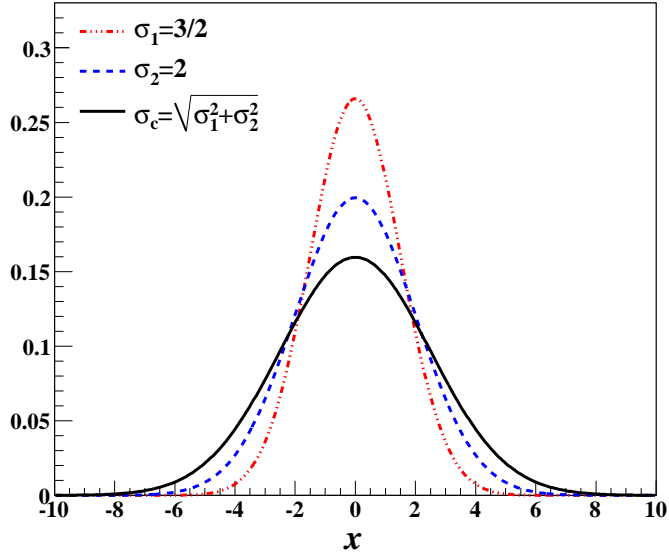}
\caption{Gaussian distributions $f(x)$, $g(x)$ and $h(x)$ used in the example of the artificial fluctuations introduced 
on histograms as a result of reusing individual showers to simulate the detector response. $f(x)$ (solid line) is the 
result of the convolution between the other two distribution functions, $g(x)$ and $h(x)$.
\label{GaussFig}}
\end{center}
\end{figure}

As an example, let us consider that $g(x)$ and $h(x)$ are two Gaussian distributions centered at zero with 
$\sigma_1=3/2$ and $\sigma_2=2$, respectively, i.e., $g(x)=G(x;0,\sigma_1)$ and $h(x)=G(x;0,\sigma_2)$, where
\begin{equation}
\label{Gauss}
G(x;\mu,\sigma) = \frac{1}{\sqrt{2 \pi}\ \sigma} \exp\left[ -\frac{(x-\mu)^2}{2\ \sigma^2}  \right]. \\
\end{equation}
The convolution of two Gaussian distributions is also a Gaussian, therefore, in this example
$f(x)$ is also a Gaussian centered at zero with $\sigma_c=[\sigma_1^2+\sigma_2^2]^{1/2}$, i.e.,
$f(x)=G(x;0;\sigma_c)$. Figure \ref{GaussFig} shows the three Gaussian distributions under consideration.

If the bin size of the histogram is sufficiently small, then
\begin{equation}
\label{IntApp}
\int_{t_k}^{t_{k+1}} dx\ A(x) \cong A(t_k) \Delta t,
\end{equation}
is a good approximation for any function $A(x)$ considered in the example. Combining this approximation with 
Eq. (\ref{Varhisnrep}) and using the Gaussian functions $f(x)$, $g(x)$ and $h(x)$, the following expression for 
the variance of $n'_k$ is obtained
\begin{eqnarray}
Var[n'_k] &=& M m\ G(t_k;0,\sigma_c) \Delta t\ (1-G(t_k;0,\sigma_c) \Delta t)+\frac{M m (m-1) \Delta t^2}{2 \sqrt{\pi}}%
\times \nonumber \\
\label{Varnex}
&& \left[ \frac{1}{\sigma_2}\ G(t_k;0,\sqrt{\sigma_1^2+\sigma_2^2/2})-\frac{1}{\sigma_c} 
\ G(t_k;0,\sigma_c/\sqrt{2}) \right].
\end{eqnarray}

\begin{figure}[!bt]
\begin{center}
\includegraphics[width=10cm]{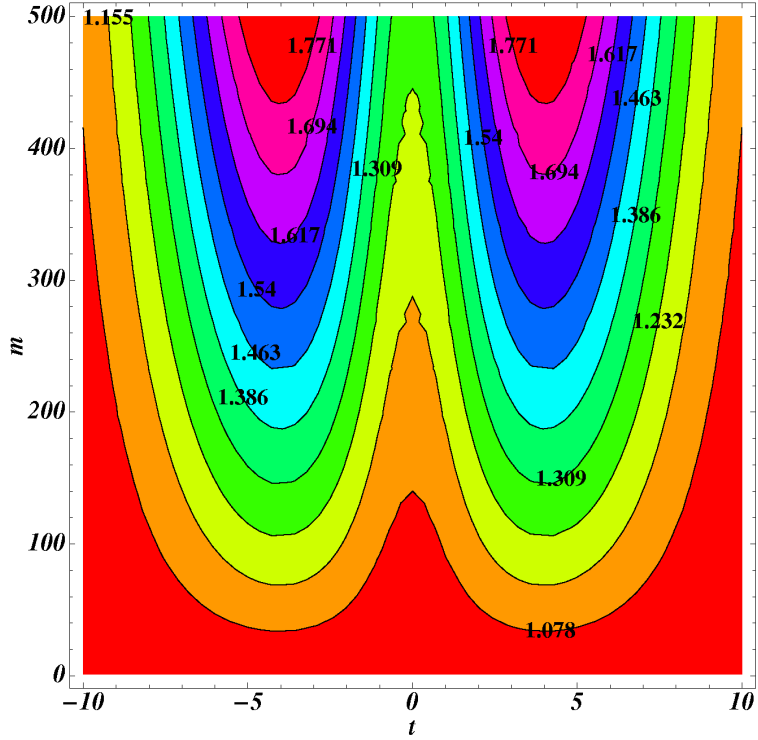}
\caption{Contour plot of $\sigma[n'_k]/\sigma[n_k]$ as a function of the lower limit of $kth$-bin $t$ and the 
number of repetitions $m$. The number of independent showers and the bin size are $M=50$ and $\Delta t=0.1$,
respectively.   
\label{HistFluct}}
\end{center}
\end{figure}

Figure \ref{HistFluct} shows a contour plot of the ratio $\sigma[n'_k]/\sigma[n_k]=Var[n'_k]^{1/2}/Var[n_k]^{1/2}$,
i.e., with  ($m>1$) and without  ($m=1$) the inclusion of shower repetitions, as a function of $m$ and $t$, the lower 
limit of $kth$ bin. The number of independent showers is taken as $M=50$ and $\Delta t=0.1$. From the figure it can 
be seen that the larger the number of repetitions the larger the fluctuations compared to the case $m=1$. 

\subsection{Density estimators}
\label{DensEst}

The density estimation technique consist in obtaining an estimator of the underlying density function from a given 
data sample \cite{Silvermann:86,Scott:92,Fadda:98,Merritt:94}. In one of the most widely used variants of that technique, 
a density estimator is obtained from a superposition of kernel functions centered at each event of the data sample. For 
$d$-dimensional data the kernel density estimator can be written as,
\begin{equation}
\label{Dest}
\hat{f}(\mathbf{x}) = \frac{1}{N}\ \sum_{\alpha = 1}^N \frac{1}{\sqrt{|H|}}\ K(H^{-1/2} \cdot (\mathbf{x}-\mathbf{x}_\alpha)),
\end{equation}
where $\mathbf{x}$ is a $d$-dimensional vector, $H$ is a symmetric, positively defined matrix (i.e., the symmetric, 
positively defined square-root matrix $H^{-1/2}$ exists) and $K(\mathbf{u})$ is the kernel function. The matrix $H$ 
gives the covariance between the different pairs of variables and also the degree of smoothing, i.e., the width of the 
kernel function.   

From Eqs. (\ref{Dsamp}) and (\ref{Dest}) the expected value of the density estimator is obtained,
\begin{equation}
\label{Bias}
E[\hat{f}(\mathbf{x})] = \frac{1}{\sqrt{|H|}}\ \int d\mathbf{x}' \ K(H^{-1/2} \cdot (\mathbf{x}-\mathbf{x}'))\ f(\mathbf{x}'),
\end{equation}
which shows that $\hat{f}(\mathbf{x})$ is a biased estimator of $f(\mathbf{x})$.

There are several criteria to measure the goodness of the density estimator. In particular the mean square error 
$MSE(\mathbf{x})=E[(\hat{f}(\mathbf{x})-f(\mathbf{x}))^2]$ is a natural criterion pointwise. Globally, $MSE(\mathbf{x})$ 
can be integrated over $\mathbf{x}$ to give the integrated mean square error,
\begin{equation}
\label{IMSE}
IMSE = \int d\mathbf{x}\ MSE(\mathbf{x}) = \int d\mathbf{x}\ E[(\hat{f}(\mathbf{x})-f(\mathbf{x}))^2]. 
\end{equation}

It is easy to see that $MSE(\mathbf{x})=Var[\hat{f}(\mathbf{x})]+Bias^2(\mathbf{x})$, where 
$Var[\hat{f}(\mathbf{x})]=E[\hat{f}(\mathbf{x})^2]-E[\hat{f}(\mathbf{x})]^2$ and 
$Bias^2(\mathbf{x})=(E[\hat{f}(\mathbf{x})]-f(\mathbf{x}))^2$. Then,
\begin{equation}
\label{IMSEVarBias}
IMSE = \int d\mathbf{x}\ Var(\mathbf{x}) + \int d\mathbf{x}\ Bias^2(\mathbf{x}). 
\end{equation}

By using the Taylor expansion and retaining the dominant terms an approximated expression for $IMSE$ is
obtained,
\begin{equation}
\label{IMSEVarBias}
IMSE \cong \frac{1}{4}\ \int d\mathbf{x}\ \left[ \int d\mathbf{u}\ K(\mathbf{u}) \mathbf{u}^T H^{1/2} D^2%
f(\mathbf{x}) H^{1/2} \mathbf{u}  \right]^2 + \frac{1}{N \sqrt{|H|}}\ R(K),
\end{equation}
where
\begin{eqnarray}
\label{D2f}
[D^2f(\mathbf{x})]_{ij} &=& \frac{\partial^2 f}{\partial x^i \partial x^j}(\mathbf{x}), \\
\label{R}
R(A) &=& \int d\mathbf{u}\ A^2(\mathbf{u}).
\end{eqnarray}

If $H^{-1/2}=V^{-1/2}/h$, where $h$ is a small parameter that parametrizes the degree of smoothing, the 
$IMSE$ is written as,
\begin{equation}
\label{IMSEVarBiash}
IMSE \cong \frac{h^4}{4}\ \int d\mathbf{x}\ \left[ \int d\mathbf{u}\ K(\mathbf{u}) \mathbf{u}^T V^{1/2} D^2%
f(\mathbf{x}) V^{1/2} \mathbf{u}  \right]^2 + \frac{R(K)}{N\ h^d\ \sqrt{|V|}}.
\end{equation}
Minimizing $IMSE$ with respect to $h$, the well known expression of $h_{opt}$ is recovered,
\begin{equation}
\label{hopt}
h_{opt} \propto \frac{1}{N^{1/(d+4)}},
\end{equation}
where the constant of proportionality depends on $f(\mathbf{x})$, the unknown density function that we want to 
estimate. There are several methods to estimate the smoothing parameter $h$ from the data sample
(see section \ref{example}).

Let us consider the case in which shower repetitions of individual showers are included. The density estimator
in this case is given by,
\begin{equation}
\label{DestRep}
\hat{f}'(\mathbf{x}) = \frac{1}{M m}\ \sum_{\alpha = 1}^M \sum_{a = 1}^m \frac{1}{\sqrt{|H|}}\ K(H^{-1/2} \cdot %
(\mathbf{x}-\mathbf{x}_{\alpha a})),
\end{equation}

It can be seen from Eqs. (\ref{DsampRep}) and (\ref{DestRep}), that the bias does not change when the repetitions are 
introduced. However, as expected, the variance increases,
\begin{eqnarray}
Var[\hat{f}'(\mathbf{x})]\ &\cong& \frac{1}{M m\ \sqrt{|H|}}\ R(K)\ f(\mathbf{x}) + \frac{m-1}{M m} \times \nonumber \\ 
\label{VarRep}
&&\left( \int d\mathbf{y}\ g(\mathbf{y}) h^2(\mathbf{x}-\mathbf{y})-f^2(\mathbf{x})\right),  
\end{eqnarray}
where just the leading terms are retained. Consequently, the $IMSE$ takes in this particular case the form 
\begin{eqnarray}
IMSE' &\cong& \frac{h^4}{4}\ \int d\mathbf{x}\ \left[ \int d\mathbf{u}\ K(\mathbf{u}) \mathbf{u}^T V^{1/2} D^2%
f(\mathbf{x}) V^{1/2} \mathbf{u}  \right]^2 + \nonumber \\
\label{IMSEVarBiashRep}
&& \frac{R(K)}{M m\ h^d\ \sqrt{|V|}}+ \frac{m-1}{M m} \left( \int d\mathbf{x} d\mathbf{y}\ g(\mathbf{y})%
h^2(\mathbf{x}-\mathbf{y}) -R(f) \right).
\end{eqnarray}

Eq. (\ref{IMSEVarBiashRep}) shows that the leading term introduced by the repetitions does not depend on $h$ and, 
therefore, the expression for $h_{opt}$ remains equal to the $m=1$ case. The only effect introduced by the repetitions 
of the individual showers is to increase the fluctuations of the estimator for each $\mathbf{x}$.

\section{Numerical Example}
\label{example}

In this section a numerical example that shows the predicted effects introduced by the shower repetitions is given. 
For that purpose, air showers simulations are performed using the program CONEX. A total of $N_{sh}=11000$ proton 
showers of primary energy $E=10^{19}$ eV and zenith angle $\theta = 30^\circ$ are generated.  

Samples of the parameter $X_{max}$ obtained from the CONEX simulations are considered. A Gaussian uncertainty of 
$\sigma[X_{max}]=20$ g cm$^{-2}$ and $\mu=0$ is assumed in order to take into account the detector response and the 
reconstruction method. Therefore, the distribution function of the reconstructed $X_{max}$ is given by Eq. (\ref{fdef}) 
with $g(X_{max})$ the distribution function corresponding to the physical fluctuations and $h(X)=G(X;0,\sigma[X_{max}])$ 
(see Eq. (\ref{Gauss})) takes into account the response of the detectors and reconstruction methods. 

Four sets of 100 samples are considered. Each set of samples is noted as $S_{(M,m)}$ where $M$ indicates the independent 
values of $X_{max}$ (obtained from CONEX) in each sample and $m$ the number of repetitions of each shower, i.e., 
the number of times that the Gaussian distribution $h(X)=G(X;X_{max}^i,\sigma[X_{max}])$ is sampled for each of the $M$ 
independent values $X_{max}^i$ in each individual sample. Therefore, $S_{(110,1)}$, $S_{(10,11)}$, $S'_{(110,1)}$ 
and $S_{(22,5)}$ are considered, where $S_{(110,1)}$ and $S'_{(110,1)}$ just differ in the different values obtained
from the Gaussian distribution performed to include the detector response and reconstruction method. The number of 
events in each sample, belonging to the different sets, is $N_{ev}=M \times m=110$, the same for all kind of samples 
considered. 
 
\begin{figure}[!bt]
\begin{center}
\includegraphics[width=6.8cm]{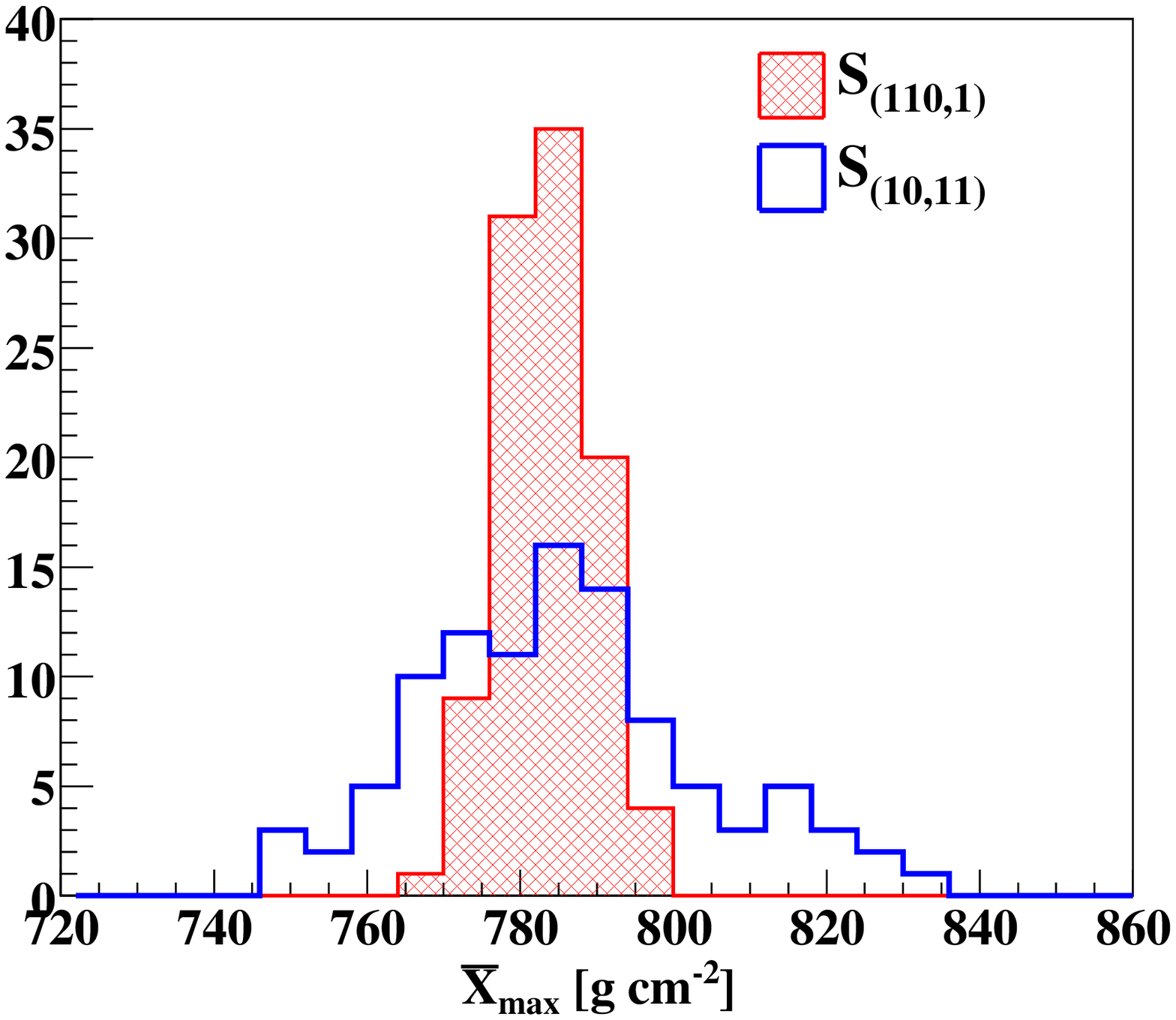}
\includegraphics[width=6.8cm]{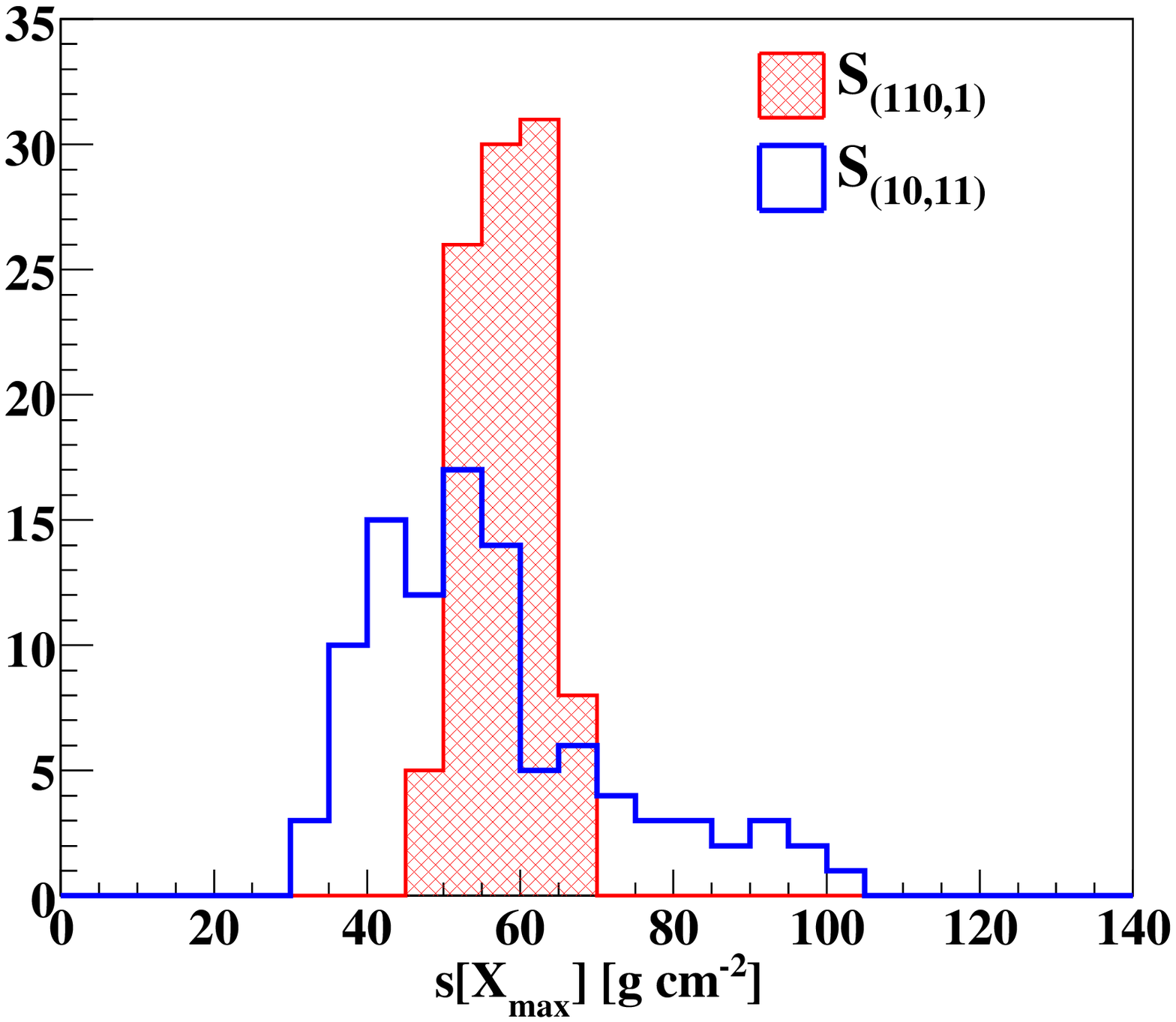}
\includegraphics[width=6.8cm]{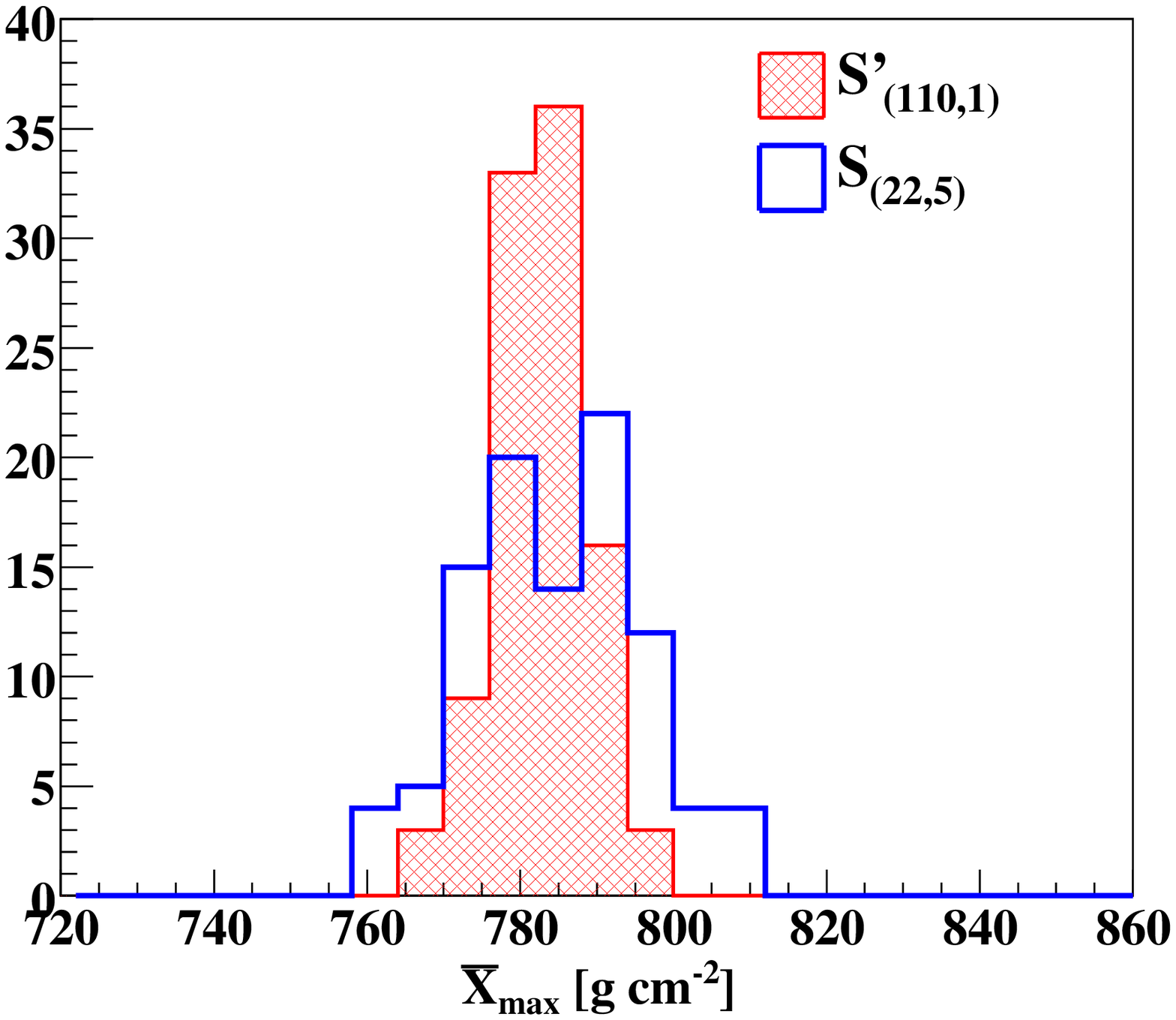}
\includegraphics[width=6.8cm]{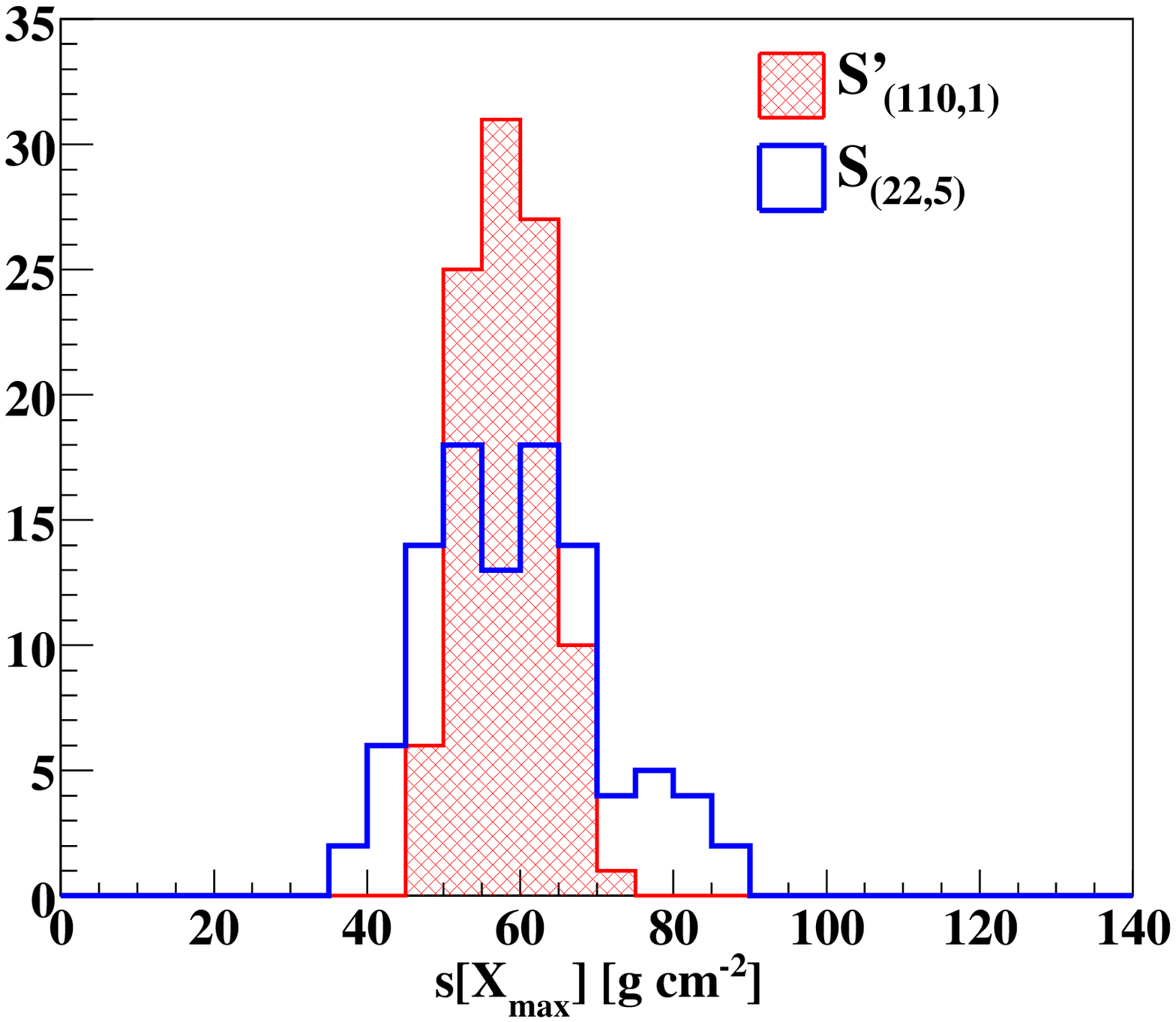}
\caption{Distributions of $\bar{X}_{max}$ and $s[X_{max}]$ for the different sets of samples considered.
Shadowed histograms correspond to samples without multiple repetitions.
\label{MeanRMS}}
\end{center}
\end{figure}

Figure \ref{MeanRMS} shows the distributions of the estimators of the average, $\bar{X}_{max}$, and the standard 
deviation, $s[X_{max}]$, for the sets of samples considered. It can be seen that, as expected, when the repetitions 
are included, the fluctuations increase and when the number of independent showers increases the fluctuations decrease.  
Figure \ref{MeanRMS} also shows that, although the distributions of $s[X_{max}]$ with repetitions have a tail towards 
larger values of grammage, which is not present in the corresponding without repetitions, the bias is not statistically 
significative. This is consistent with Eq. (\ref{Evarxrep2}) which shows that the expected bias introduced by repetitions 
in the variance is proportional to $(m-1)/M m \cong 0.1$ for $S_{(10,11)}$.     

In order to illustrate the effects of repetitions on the density estimators, one-dimensional Gaussian kernels 
are used to estimate the density function of $X_{max}$. An adaptive bandwidth method, introduced by B. Silverman 
\cite{Silvermann:86}, is used to obtain better estimates of the density function. The procedure starts by performing 
a first estimation of the density function, from a given sample, using a Gaussian kernel with fixed smoothing parameter,
\begin{equation}
\hat{P}_{0}(X_{max}) = \frac{1}{N \sqrt{2 \pi}\ \sigma\ h_{0}} \sum_{i=1}^{N}\ %
\exp\left[ -\frac{(X_{max}-X_{max}^i)^2}{2\ h^2_{0}\ \sigma^2} \right],
\label{fest0}
\end{equation}
where $N$ is the size of the sample, $\sigma$ is the standard deviation of the data sample and 
$h_{0} = 1.06 \times N^{-1/5}$ is the smoothing parameter corresponding to Gaussian
samples which is used very often in the literature because it gives very good estimates even for non
Gaussian samples.

The following parameters are calculated by using the estimate obtained from Eq. (\ref{fest0}),
\begin{equation}
\lambda_{i} = \left[ \frac{\hat{P}_{0}(X_{max}^i)}{\left( \prod_{k=1}^{N} \hat{P}_{0}(X_{max}^k)%
\right)^{1/N} } \right]^{-1/2},
\label{lamda}
\end{equation}
and then, the final density estimate is obtained from,
\begin{equation}
\hat{P}(X_{max}) = \frac{1}{N \sqrt{2 \pi}\ \sigma} \sum_{i=1}^{N}\ \frac{1}{h_{i}}%
\exp\left[ -\frac{(X_{max}-X_{max}^i)^2}{2\ h^2_{i}\ \sigma^2} \right],
\label{fest}
\end{equation}
where $h_{i} = h_{0} \ \lambda_{i}$.

\begin{figure}[!bt]
\begin{center}
\includegraphics[width=6.8cm]{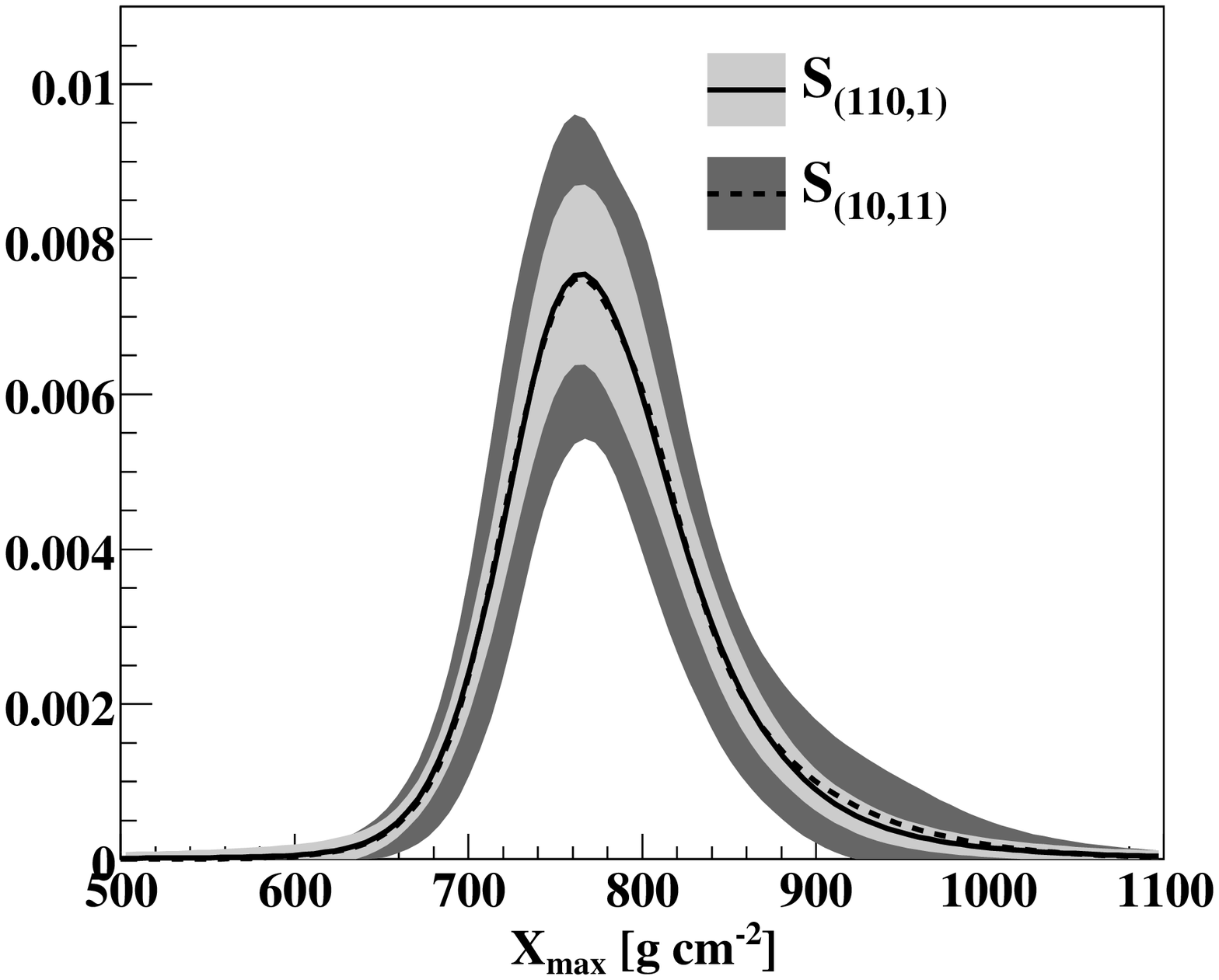}
\includegraphics[width=6.8cm]{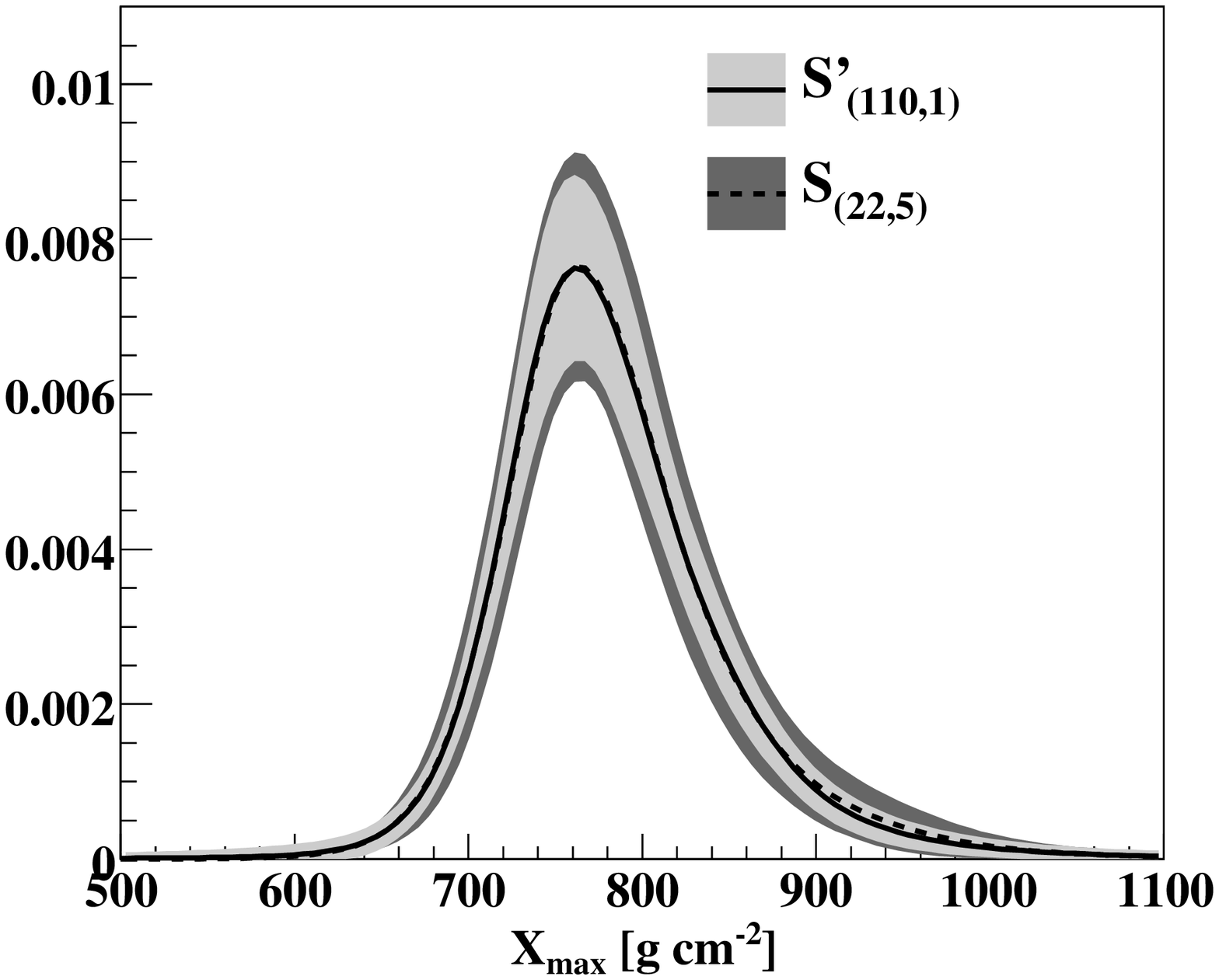}
\caption{Mean and one sigma regions for the density estimates obtained from the different samples considered. Darker
regions and dotted lines correspond to samples including multiples repetitions. 
\label{Estimates}}
\end{center}
\end{figure}

For each sample belonging to a given set a density estimate is obtained, therefore, 110 density estimates are
obtained for each set of samples considered. Figure \ref{Estimates} shows the mean value and the one sigma region 
obtained from the density estimates of each set. It can be seen that the mean values corresponding to samples with 
or without repetitions are very similar, which is consistent with the result obtained in subsection \ref{DensEst}. 
Also, as expected from Eq. (\ref{VarRep}), the fluctuations corresponding to sets including repetition are larger
and comparing the results obtained for $S_{(10,11)}$ and $S_{(22,5)}$ we see that the fluctuations in the latter 
case are smaller due to the smaller number of repetitions.  

\section{Conclusions}

In this work we study the effects of recycling individual cosmic ray showers to simulate the detector response,
which is a common practice in Monte Carlo simulations at the highest energies. We find that the standard 
estimators of the expected value, variance and covariance are modified. In particular, the average remains as a 
non-biased estimator of the expected value but the fluctuations are increased. For the standard estimators of the 
variance and covariance a bias proportional to $(m-1)/Mm$ appears when repetitions are included. 
Besides, as in the case of the average, the fluctuations of both estimators are increased. We also study the effects 
of repetitions in histograms, where we find that the mean value of the bin content is unchanged but the fluctuations are
in general larger, depending on the bin considered, and increase with the number of repetitions.    

Finally, we study the effects introduced by repetition in the kernel density estimators obtained from finite
samples. We find again that the expected value of the estimator is unchanged, i.e., the bias takes the same form. 
However, the pointwise fluctuations are increased and become more important as the ratio $(m-1)/Mm$ increases.

\section{Acknowledgments}

The authors acknowledge the support of UNAM through PAPIIT grant IN115707 and CONACyT through its research grants 
and SNI programs. ADS is supported by a postdoctoral grant from the UNAM.


\begin{thebibliography}{00}

\bibitem{Alan:00} M. Nagano and A. A. Watson, Rev. Mod. Phys. {\bf 72}, 689 (2000).

\bibitem{AIRES} S. Sciutto, AIRES user's Manual and Reference Guide (2002),
http://www.fisica.unlp.edu.ar/auger/aires.

\bibitem{CORSIKA} D. Heck {\em et al.}, Report {\bf FZKA 6097}, Forschungszentrum
Karlsrue, 1998; http://www-ik3.fzk.de/$\sim$/heck/corsika.

\bibitem{conex} T. Bergmann et. al., Astropart. Phys. {\bf 26}, 420 (2007) and T. Pierog et. al., Nucl. Phys. Proc. 
Suppl. {\bf 151}, 159 (2006).

\bibitem{Hillas:85} A. Hillas, Proc. 19th ICRC {\bf 1}, 155 (1985).

\bibitem{Hillas:97} A. Hillas, Nucl. Phys. (Proc. Suppl.) {\bf B52}, 29 (1997).

\bibitem{Ave:01} M. Ave J. Knapp, J. Lloyd-Evans, M. Marchesini and A. Watson, 
Astropart. Phys. {\bf 19}, 47 (2003).

\bibitem{Ave:02} M. Ave et. al., Astropart. Phys. {\bf 19}, 61 (2003).

\bibitem{Dova:03} M.T. Dova, M.E. Mancenido, A.G. Mariazzi, T.P. McCauley and A.A. Watson, 
Astropart. Phys. {\bf 21} 597 (2004).

\bibitem{DovaICRC:03} M. Dova, M. Mancenido, A. Mariazzi, T. McCauley and A. Watson,
Proceedings of 28th International Cosmic Ray Conferences, Tsukuba, Japan,
377 (2003).

\bibitem{Vitor:05} V. de Souza, G. Medina-Tanco and J. Ortiz, Phys. Rev. {\bf D72} 103009 (2005).  

\bibitem{VitorFede:05} V. de Souza, G. Medina-Tanco, J. Ortiz and F. Sanchez, Phys.Rev. {\bf D73} 
043001 (2006).

\bibitem{Rebel:95} H. Rebel, G. V\"olker, M. F\"oller and A. Chilingarian, J. Phys. G: Nucl. 
Part. Phys. {\bf 21}, 451 (1995).

\bibitem{Brancus:97} I. Brancus et. al., Astropart. Phys. {\bf 7}, 343 (1997).

\bibitem{Antoni:07} T. Antoni et. al., Astropart. Phys. {\bf 16}, 245 (2002).

\bibitem{Antoni:03} T. Antoni et. al., Astropart. Phys. {\bf 18}, 319 (2003).

\bibitem{Brancus:03} I. Brancus et. al., J. Phys. {\bf G29}, 453 (2003).

\bibitem{SupaCompo:08} A. D. Supanitsky, G. Medina-Tanco and A. Etchegoyen, submitted to 
Astropart. Phys. (2008).

\bibitem{Silvermann:86} B. Silvermann, \emph{Density Estimation for Statististics and Data Analysis},
ed. Chapman \& Hall, New York (1986).

\bibitem{Scott:92} D. Scott, \emph{Multivariate Density Estimation}, ed. Wiley, New York (1992).

\bibitem{Fadda:98} D. Fadda, E. Slezak y A. Bijaoui, Astron. Astrophys. Suppl. Ser. {\bf 127}, 335 (1998).

\bibitem{Merritt:94} D. Marritt y B. Tremblay, Astron. J. {\bf 108}, 514 (1994).




\end{thebibliography}
\end{document}